# Synthesis of ammonium silicon fluoride cryptocrystals on silicon by dry etching

**Seref Kalem**

TUBITAK – UEKAE, The Scientific and Technical Research Council of Turkey – National Institute of Electronics and Cryptology, Gebze 41470 Kocaeli, TURKEY

E-mail address:  s.kalem@uekae.tubitak.gov.tr

**Abstract:**

Cryptocrystal layers of ammonium silicon fluoride $(NH_4)_2SiF_6$ were synthesized on silicon wafers by dry etching method using vapor of the mixture of HF and $HNO_3$ solutions at room temperature.  Crystalline layers having thicknesses of up to 8μm have been produced at growth rates of around 1 μm/hour.  The crystallinity was analysed by X-ray diffraction that indicates an isometric hexoctahedral system (4/m -32/m) with Fm3m space grouping of $(NH_4)_2SiF_6$ cryptohalite crystals.  These results have been confirmed by the presence of vibrational absorption bands of $(NH_4)_2SiF_6$ species by Fourier transform infrared (FTIR) spectroscopic measurements.  Strong absorption bands were observed in the infrared at 480cm$^{-1}$, 725cm$^{-1}$, 1433cm$^{-1}$ and 3327cm$^{-1}$ and assigned to N-H and Si-F related vibrational modes of $(NH_4)_2SiF_6$. Annealing above 150$^{o}$C leads to the formation of individual crystals with sizes up to 20μm on the surface, thus indicating the posibility of forming solid compound layers with fine grain sizes on silicon.

# 1  Introduction

Crystal texture for some materials is so finely grained that no distinct particles are discerned, even under polarization microscope. State of matter arranged in this way with such minute crystals is said to be cryptocrystalline or cryptogranular. This type of crystals can exhibit extraordinary dielectric properties which can be used in various fields ranging from microelectronics and packaging applications to photonics and optics [1,2]. The synthesis of cryptocrystals of Ammonium Silicon Fluoride $(NH_4)_2SiF_6$ (hereafter ASF) on silicon may be an important step forward for the integration of a variety of materials into silicon circuits technology. Such layers can be used as a buffer, bridging the gap between the large lattice mismatches between Silicon (Si) and other materials [3,4]. With the fluoride buffer layers, advanced semiconductor hetero-structures can be directly grown on Si wafers. Thus, inexpensive electronic and photonic devices can be produced and integrated into Si circuits technology. An added benefit resulting from the water solubility of fluorides is that the films grown on this buffer layer can be readily lifted-off for some practical applications, such as heat sinking and regrowth. The heat sinking performed in this way may allow laser operation at high temperature.

The ASF crystals have been generated as by-products during the silicon native oxide cleaning experiments using different processes and chemistry [5]. However, there has not been any study on the synthesis of ASF cryptocrystals on silicon. In these experiments, a process involving hot $NH_3/NF_3$ mixture has been used to remove Si native oxide from contact holes. This process has generated ASF on the surface as a result of the reaction of $NH_3$ with $NF_x$ (x=1,2) species at high temperature ($600^oC$).

The dry etching technique used in this work and described elsewhere [6] can be used for direct formation of advanced lithographic structures on silicon. In addition to Si, other potential substrates such as $SiO_2$, $Si_3N_4$ and SiGe can be processed using this technique. The resulting layers from this process can be used as etch stop and device isolation layers or diffusion barrier in metallization applications. Moreover, the process itself can provide further insight into the understanding of porous structure formation mechanism in silicon [7,8], since the same chemicals are involved in the material processing.

In this paper, we demonstrate the possibility of growing fluoride layers by a dry etching technique and investigate the physical properties of the resulting material. Formation of such layers offers the possibility of writing on substrates by selectively exposing Si surface to the vapors of HF and $HNO_3$ acid mixture. Growing layers by dry etching has several advantages: i) no electrical contacts are needed; ii) selective processing is possible; iii) layers are homogeneous; iV) thickness can be controlled; v) they can serve as diffusion barrier; vi) offers a cost-effective solution to buffer layer growth compared to other techniques, such as molecular beam epitaxy, chemical vapor deposition and even spin-on technology.

## 2 Experimental

Cryptocrystalline thin films of Ammonium Silicon Fluoride $(NH_4)_2SiF_6$ (ASF) have been synthesized on Si wafers during our attempts to grow porous silicon layers from the vapors of chemical etchants [6]. The fluoride cryptocrystals were formed when Si surface was exposed to the vapors of a mixture of hydrofluoric acid (HF) and nitric acid ($HNO_3$). The HF:$HNO_3$ mixture is known as stain-etching solution in the porous silicon formation process [9,10]. The volume ratio of the solutions of HF:$HNO_3$ mixture was between 7:1 – 7:3.6 for the sample preparation. The chemicals used in the process were semiconductor grade 40% HF and 65% $HNO_3$ by weight. The samples in these experiments were made at room temperature using Boron-doped p-type (100) Si wafers with resistivities of 5-10 Ohm-cm. But, the ASF crystals were also successfully grown on both the silicon nitride ($Si_3N_4$), SiGe and the n-type Si wafers with {100} and {111} crystalline orientations. It was recently shown that the porous silicon layers can be formed on both the p-type and n-type Si wafers from the vapors of aqueous HF etching solutions[6]. The details of experimental setup for the growth of porous silicon layers from the vapor phase etchants were reported earlier [7].

The layers of ammonium silicon fluoride are probably formed by silicon mediated coupling reactions between HF and $HNO_3$ species on the wafer (Si) surface. Unfortunately, description of detailed chemical reaction leading to the synthesis of ammonium silicon fluoride is not yet known. The reason is mainly due to the lack of evidence for a complete account of reaction products. We think there is not only ammonium silicon fluoride, but also oxygen $O_2$ comes out as a reaction product. Nevertheless, this assumption has to be proved experimentally. After the reaction has been stopped, the fluoride layer was removed from the surface for x-ray analysis. The layer can be removed also by thermal decomposition at $150^oC$ or by dissolution in water. Both the processes result in selective bulk etching of Si wafer. It was found that the same dry etching process has also been effective for the fluoride growth on both the $SiO_2$ , SiGe and $Si_3N_4$ substrates.

Crystal structure, vibrational and surface properties of the layers were determined using X-ray diffraction (XRD) analysis, Fourier transform infrared (FTIR) spectroscopic measurements and scanning electron microscopy (SEM) , respectively. The as-grown layers on Si surface are smooth and exhibit a white powder like appearance, which are readily soluble in water. The presence of interference fringes in the FTIR spectra also indicates a homogeneous film with reasonable optical quality. The FTIR spectra revealed all the related N-H and Si-F vibrational modes in the fluoride layer. The crystal structure were determined using X-ray diffraction measurements on free-standing layers. The annealing experiments were carried out between $50\text{-}175^oC$ and enabled us to study structural integrity and composition of the material.

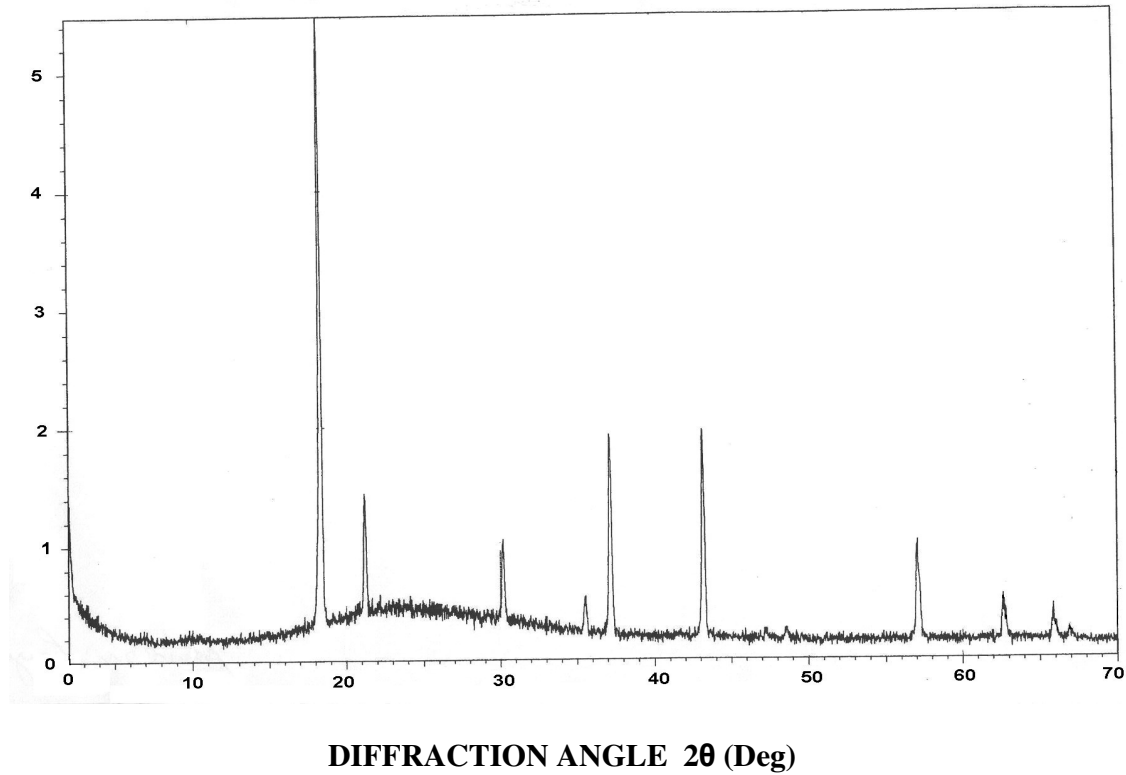

**Figure 1.** X-ray diffraction pattern of free-standing cryptocrystal layers of ammonium silicon fluoride $(NH_4)_2SiF_6$ grown by dry etching in vapor of HF and $HNO_3$ solutions with a ratio of 3:1. Sharp peaks arise from the randomly oriented fluoride cryptocrystals in the film.

## 3  Results and discussion

The crystallographic structure was investigated by X-ray diffraction measurements using Cu $K_\alpha$ radiation. The sharp peaks indicate that the film has grown in crystalline structure with preferred orientation of {111}. Figure 1 shows the x-ray diffraction intensity as counts per second taken from our free standing samples at room temperature. The strongest 3 peaks are located at $2\theta$ =18.34 deg, 43.14 deg and 37.14 deg, where $\theta$ is the diffraction angle. Corresponding interlayer spacing-d is estimated using Braggs law ( $n\lambda=2d\sin\theta$, where $\lambda=1.54$Å and n=1) to be 4.834 Å, 2.096 Å and 2.419 Å, respectively. The major diffraction peaks of the ASF layers are summarized in Table 1. The results of these analysis are in agreement with the data reported on cryptohalite crystals of ammonium silicon fluoride $(NH_4)_2SiF_6$ with a lattice constant of 8.395Å [11]. These results suggest crystalline character of the layer which is the cryptocrystals of ammonium silicon fluoride $(NH_4)_2SiF_6$ with isometric hexoctahedral symmetry (4/m -32/m) with space group Fm3m.

**Table 1** Major diffraction peak data for the ammonium silicon hexafluoride crystalline films. $I/I_1$ is the normalized intensity.

| Peak No. | 2 Theta (Degree) | d (Å) | $I/I_1$ |
| --- | --- | --- | --- |
| 1 | 18.3401 | 4.83355 | 100 |
| 2 | 21.2009 | 4.18734 | 19 |
| 3 | 30.1452 | 2.96221 | 15 |
| 4 | 35.4952 | 2.52703 | 7 |
| 5 | 37.1360 | 2.41906 | 39 |
| 6 | 43.1362 | 2.09545 | 43 |
| 7 | 57.0333 | 1.61348 | 22 |
| 8 | 62.6247 | 1.48219 | 9 |
| 9 | 65.8394 | 1.41739 | 7 |

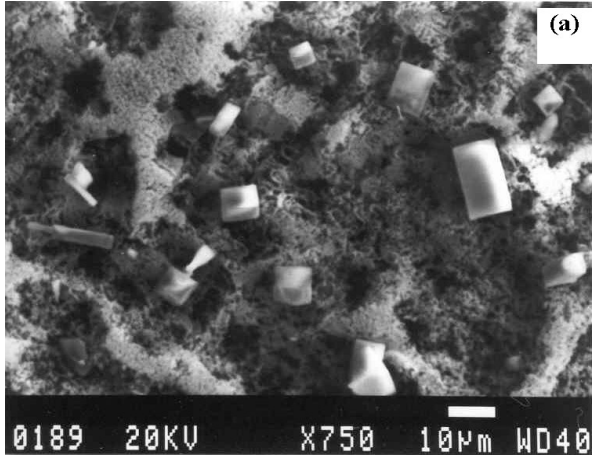

**Figure 2.** **a)** SEM micrograph (at X750 magnification) of ammonium silicon fluoride $(NH_4)_2SiF_6$ cryptocrystals after annealing at 175 °C. The surface is decomposed and the formation of individual fluoride crystals are readily seen on the image. **b)** Further magnification (X900) reveals the crystals of different sizes and shapes. The thermal annealing was carried out at atmospheric pressure and without surface protection against decomposition. Note that the upper left hand side of the micrograph shows that there is still a layer sticking on the surface.

The as-grown layers are in the form of crystalline white granular, porous film with a smooth surface. In order to identify the nature of the layer, thermal annealing experiments were carried out on a hot plate. During this process, the surface is exposed to air but not protected against possible decomposition and outgassing. We observed that the surface integrity was kept intact with thermal annealing up to about 150 °C. But, the decomposition of the surface with high density of pits starts to occur at temperatures greater than 150°C. Also, at higher temperature, we observed the formation of individual fluoride crystals scattered throughout the surface of the wafer as shown in SEM micrograph in Fig. 2. In this particular case, large crystals of up to about 20 μm can be obtained on the samples. Also, the formation of deep micropores was observed on the Si surface. Note that some parts of the surface were not decomposed, exhibiting a solid compound like structure.

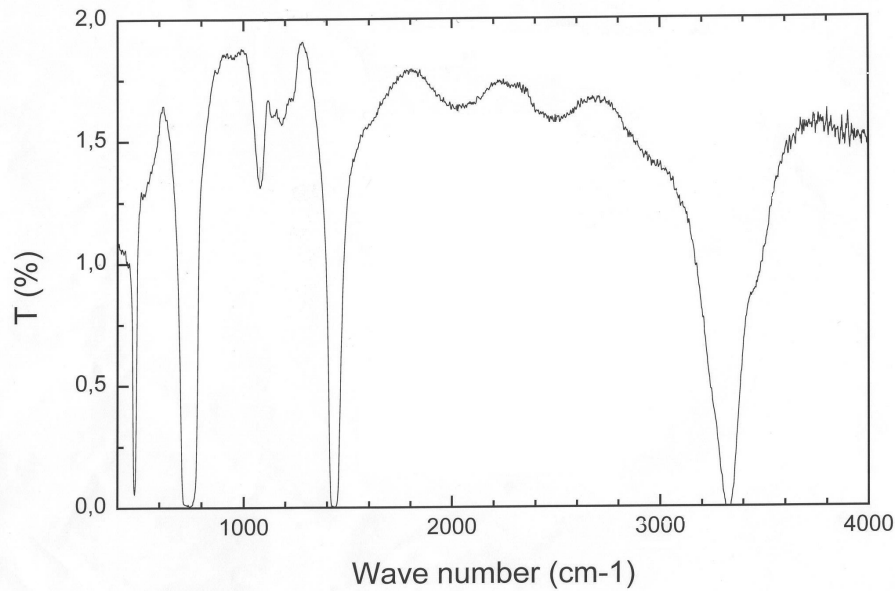

**Figure 3.** Room temperature FTIR transmission spectrum of as-grown ammonium silicon fluoride $(NH_4)_2SiF_6$ layer on Si. Vibrational bands are of N-H and Si-F stretching modes, typical of $(NH_4)_2SiF_6$ species.

Optical properties of the layers were analyzed by Fourier transform infrared (FTIR) spectroscopic measurements between 400 cm$^{-1}$ and 4000 cm$^{-1}$. Figure 3 shows the typical room temperature FTIR transmission spectrum of as-grown fluoride layer . From the interference fringes in the IR spectrum, we estimated a thickness of 8.1μm using a refractive index of n = 1.369 for the ammonium silicon fluoride. The spectrum exhibits a large number of strong IR absorption bands, in addition to usual but, relatively weak Si-O stretching mode at 1083 cm$^{-1}$ with a shoulder at 1180 cm$^{-1}$. We observed very strong vibrations at 480 cm$^{-1}$, 1433 cm$^{-1}$ , 725

cm$^{-1}$ (this mode is a doublet with peaks at 715cm$^{-1}$ and 740cm$^{-1}$), and an asymmetric band at 3327 cm$^{-1}$ with a shoulder at 3449 cm$^{-1}$. Ammonium silicon fluoride (NH$_4$)$_2$SiF$_6$ (99.9999%) as an inorganic material exhibits vibrational modes at 480 cm$^{-1}$, 725 cm$^{-1}$, 1433 cm$^{-1}$ and 3327 cm$^{-1}$ [12]. The frequencies of these vibrations are in a very good agreement with the IR vibrational bands observed in our samples. By analogy to previous studies, we have assigned these bands to various N-H and Si-F related vibrational modes and the results are shown in Table 2. Note that the Si-O vibrations are originated from the oxide structure at the interface between the layer and Si substrate.

**Table 2**   Summary of FTIR data for (NH$_4$)$_2$SiF$_6$ cryptocrystals grown on Si. Observed vibrational frequencies, their tentative assignments and absorption intensities are listed in the first, second and third columns, respectively.   Relative peak intensities are indicated as VS: Very strong, S: Strong, M: Medium, W: Weak, VW: Very weak.

| Peak frequencies $\omega$(cm$^{-1}$) | Tentative assignment | Intensity |
|---|---|---|
| 480 | N-H wag. or Si-F def. | VS |
| 725 | Probably Si-F or Si-N stretching | VS |
| 1083 | Si-O stretching | M |
| 1180 | Si-O asym. Str. | W |
| 1433 | N-H bending or deformation | VS |
| 2125 | Si-H stretching | VW |
| 3327 | N-H symmetric stretching | VS |
| 3449 | N-H degenerate stretching | M |

In order to assign the IR absorption bands, the layers were annealed between 50 °C and 150°C on a hot plate under atmospheric pressure without any surface protection. The annealing of the sample at around 150 °C resulted in narrowing of all the band widths as shown in Fig.4. We observed a narrowing by about 67% in FWHM of the N-H stretching band at 3327cm$^{-1}$. The asymmetric nature of the band was disappeared due to weaker strength of the shoulder. The difference in the IR-peak widths are probably due to grain size with packing density increased indicating that the annealed sample is uniformly very fine grained. The smearing out of the interference fringes and the change at the position of their maxima and minima in the IR transmission spectra are probably due to both the surface decomposition and rearrangement of grains resulting in a subsequent decrease in the layer thickness. After annealing, the thickness was estimated to be 7.7μm from the remaining fringes, thus indicating a decrease of 0.4μm (or 5% decrease) at the thickness. The formation of individual large crystals and the complete surface decomposition occur at much higher temperatures as seen on the SEM micrographs in Fig. 2.

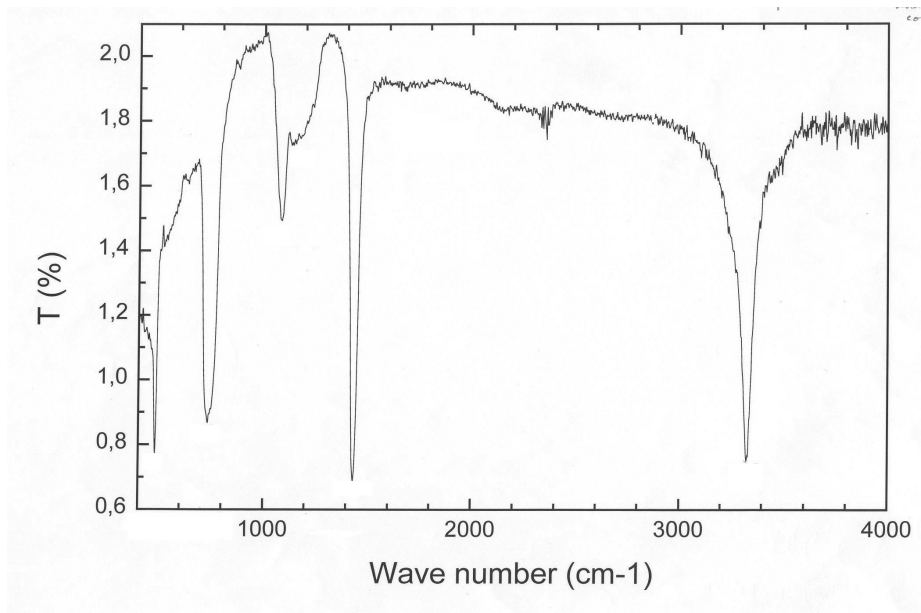

**Figure 4.** FTIR transmission spectrum of ammonium silicon fluoride $(NH_4)_2SiF_6$ cryptocrystals on Si substrate after thermal annealing at 150 °C.

Another feature of the IR vibrational bands is related to the behavior of Si-O stretching modes at 1083cm$^{-1}$ and 1180cm$^{-1}$, indicating the importance of surface oxidation and thus the strain build-up at the interface. We observed that these bands increased by 35% on the effect of the thermal annealing. The intensity of the oxygen bands also increased with exposure of the layer to air for an extended period of time. This property suggests that surface should be protected against oxidation in order to avoid strain related problems at the interface. The presence of very weak Si-H bands at 635cm$^{-1}$ and 2125cm$^{-1}$ in the spectrum of the annealed sample suggests that some hydrogenation effect also takes place with thermal annealing. In this case, the hydrogen probably originates from the HF molecules released from the thermal decomposition of $(NH_4)_2SiF_6$, thus leading to hydrogen terminated surface.

## 4  Conclusion

Cryptocrystal layers of ammonium silicon fluoride $(NH_4)_2SiF_6$ were successfully grown on Si wafers by exposing the surface to the vapor of the mixture of HF and $HNO_3$ solutions at room temperature. The crystalline structure of these layers were attributed to isometric hexoctahedral symmetry of ammonium silicon fluoride $(NH_4)_2SiF_6$ cryptohalite crystals. Vibrational spectra confirmed the XRD results and revealed several useful notches at 3μm, 7μm, 13.6μm and 20.8μm for possible applications in photonics. The as-grown layers exhibit porous, granular structure and thermal annealing leads to uniformly very fine grained layers. The formation of these low-**k** (**k**=1.87) dielectric layers by this simple and cost effective dry etching technique may find important applications in the fields ranging from device fabrication to photonics. There are lots of challenges to address with this technology as well as new opportunities for further studies.

**Acknowledgment**

Author is grateful to Dr. E. Gunay and Z. Misirli for providing the SEM picture and x-ray diffraction curve.